\documentstyle[preprint,aps]{revtex}

\begin{document}
\draft
\preprint{TUHEP-TH-02133}
\title{Fermion determinant with dynamical chiral symmetry breaking}

\author{Qin Lu$^a$,  Hua Yang$^{a,b}$, Qing Wang$^{a,c}$}

\address{$^a$Department of Physics, Tsinghua University, Beijing 100084, China
\footnote{Mailing address}\\
$^b$Institute of Electronic Technology, Information Engineering University, Zhengzhou 450004\\
$^c$Institute of Theoretical Physics, Academia Sinica, Beijing 100080, China}
\date{Feb 1, 2002}

\maketitle
\begin{abstract}
One -loop fermion determinant is discussed
for the case in which dynamical chiral symmetry breaking caused by momentum
dependent fermion self energy $\Sigma(p^2)$ take place. The obtained series
generalizes the heat kernel  expansion for hard fermion mass.
\end{abstract}

\bigskip
PACS number(s): 03.65.Db, 11.10Ef, 11.30.Rd, 12.39.Fe

\vspace{1cm}

In quantum field theory, 1-loop quantum correction of fermions is
from fermion determinant. It is known how to derive an exact
non-perturbative representation for the
 chiral fermion determinant with hard fermion mass\cite{Ball} and the technique has been widely used in the literature  such as to derive the  effective action of quarks aiming at modeling QCD at low-energy \cite{ENJL,Espiru,WQ}.
The central object in the calculation of the effective action is lndet$(D+m)$,
 where in the Euclidean space, the differential Dirac operator
 $D$ depends on hermitian external flavor sources (or collective meson fields) which
 have well defined
transformation laws with respect to the action of the chiral group
\begin{eqnarray}
&&D\equiv\nabla\!\!\!\! /\;-s+ip\gamma_5\hspace{2cm}
\nabla_{\mu}\;\equiv \partial_{\mu}-iv_{\mu}-ia_{\mu}\gamma_5
=-\nabla_{\mu}^{\dagger}\;\;,\label{nabladef}
\end{eqnarray}
$m$ is a hard fermion mass. Ignoring anomaly, the real part of lndet$(D+m)$ in
 terms of a proper time integral is
\begin{eqnarray}
{\rm Re}\ln{\rm Det}(D+m)&=&\frac{1}{2}
{\rm Tr}\ln[(D^{\dagger}+m)(D+m)]\nonumber\\
&=&-\frac{1}{2}\lim_{\Lambda\rightarrow\infty}
\int d^4x\int_{\frac{1}{\Lambda^2}}^{\infty}
\frac{d\tau}{\tau}~{\rm Tr}~e^{-m^2\tau}
\langle x|e^{-\tau(E-\nabla^2)}|x\rangle\label{lndetcal}
\end{eqnarray}
with
\begin{eqnarray}
E-\nabla^2=D^{\dagger}D+D^{\dagger}m+mD
\end{eqnarray}
and
\begin{eqnarray}
&&(\nabla\!\!\!\! /\;)^{\dagger}\equiv -\partial\!\!\!/\;+iv\!\!\!
/\; -ia\!\!\! /\;\gamma_5\hspace{3cm} D^{\dagger}=\nabla\!\!\!\!
/\;^{\dagger}-s-ip\gamma_5\nonumber\\ &&E= -2ms-2ima\!\!\!
/\;\gamma_5 +\frac{i}{4}[\gamma^{\mu},\gamma^{\nu}]{\cal
R}_{\mu\nu} +\gamma_{\mu}d^{\mu}(s+ip\gamma_5)
+i\gamma^{\mu}[a_{\mu}\gamma_5(s+ip\gamma_5)+(s+ip\gamma_5)a_{\mu}\gamma_5]
\nonumber\\ &&\hspace{1cm}+s^2+p^2+[s,p]i\gamma_5\\
 &&{\cal R}_{\mu\nu}\equiv i[\nabla_{\mu},\nabla_{\nu}]
=[d_{\mu}a_{\nu}-d_{\nu}a_{\mu}]\gamma_5+V_{\mu\nu}
-i[a_{\mu},a_{\nu}]\hspace{2cm}V_{\mu\nu}
=\partial_{\mu}v_{\nu}-\partial_{\nu}v_{\mu}-i[v_{\mu},v_{\nu}]
\nonumber\\ &&d_{\mu}f=\partial_{\mu}f-i[v_{\mu},f]\hspace{2cm}
d_{\mu}(fg)=(d_{\mu}f)g+f(d_{\mu}g)\;\;.\nonumber
\end{eqnarray}
With help of standard Seely-DeWitt expansion,
\begin{eqnarray}
\langle x|e^{-\tau(E-\nabla^2)}|x\rangle\label{Eexp}
&=&\frac{1}{16\pi^2}\bigg[\frac{1}{\tau^2}-\frac{E}{\tau}+\frac{1}{2}E^2
-\frac{1}{6}[\nabla_{\mu},[\nabla^{\mu},E]]
-\frac{1}{12}{\cal R}_{\mu\nu} {\cal R}^{\mu\nu}
-\frac{\tau}{6}E^3\label{Seely}\\
&&+\frac{\tau}{12}\{E[\nabla^{\mu},[\nabla_{\mu},E]]
+[\nabla^{\mu},[\nabla_{\mu},E]]E
+[\nabla^{\mu},E][\nabla_{\mu},E]\}+\frac{\tau^2}{24}E^4+\cdots\bigg]\;,\nonumber
\end{eqnarray}
the integration of $\tau$ in (\ref{lndetcal}) can be finished, we
then get the expansion of Re~lndet$(D+m)$ with hard fermion mass $m$,
\begin{eqnarray}
&&{\rm Re}\ln{\rm Det}(D\!\!\!\! /\;+m)\nonumber\\ &&=\ln{\rm
Det}(\partial\!\!\!\! /\;+m)
-\frac{N_c}{32\pi^2}\lim_{\Lambda\rightarrow\infty} \int d^4x{\rm
tr}_f \bigg[8m[\Lambda^2+m^2(\ln\frac{m^2}{\Lambda^2}+\gamma-1)]s
-8m^2(\ln\frac{m^2}{\Lambda^2}+\gamma)a^2\nonumber\\
&&\hspace{0.5cm}-\frac{4}{3}[d_{\mu}a^{\mu}]^2
-\frac{2}{3}(\ln\frac{m^2}{\Lambda^2}+\gamma+1)
(d_{\mu}a_{\nu}-d_{\nu}a_{\mu})(d^{\mu}a^{\nu}-d^{\nu}a^{\mu})
-[\frac{4}{3}(\ln\frac{m^2}{\Lambda^2}+\gamma)+\frac{16}{3}]a^4
\nonumber\\ &&\hspace{0.5cm}
 +[\frac{4}{3}(\ln\frac{m^2}{\Lambda^2}+\gamma)
+\frac{8}{3}]a_{\mu}a_{\nu}a^{\mu}a^{\nu}
-4[\Lambda^2+m^2(3\ln\frac{m^2}{\Lambda^2}+3\gamma-1)]s^2
\nonumber\\ &&\hspace{0.5cm}
-4[\Lambda^2+m^2(\ln\frac{m^2}{\Lambda^2}+\gamma-1)]p^2
+(16\ln\frac{m^2}{\Lambda^2}+16\gamma+16)msa^2
-\frac{2}{3}(\ln\frac{m^2}{\Lambda^2}+\gamma)V_{\mu\nu}V^{\mu\nu}
\nonumber\\ &&\hspace{0.5cm}
+i[\frac{8}{3}(\ln\frac{m^2}{\Lambda^2}+\gamma)+\frac{16}{3}]a^{\mu}a^{\nu}V_{\mu\nu}
+8m(\ln\frac{m^2}{\Lambda^2}+\gamma)pd^{\mu}a_{\mu}+O(p^6)\bigg]\;\;,\label{exp}
\end{eqnarray}
where $N_c$ is number of color (which in present discussion is a
global degree of freedom) and tr$_f$ is trace for flavor indices.
The expansion terms are arranged according to its momentum power,
$\partial_{\mu}$, $v_{\mu}$, $a_{\mu}$ are treated as order $p$
and $s$, $p$  as order $p^2$.

 The hard fermion mass $m$  play the role of breaking chiral symmetry and
infrared cutoff. It can be generalized  to a matrix \cite{matrix} to respect
 further breaking of the chiral symmetry. But the formulation  cannot deal with
 the case of dynamical symmetry breaking in which the most general situation is
 not the appearance of a hard fermion mass, but a momentum dependent fermion
  self energy  $\Sigma(p^2)$. Even in NJL \cite{NJL}  or ENJL \cite{ENJL}
  models, the hard
 mass will be replaced by momentum dependent self energy in
 high order loop calculations. Taking hard fermion mass often cause extra
 ultraviolet divergence. For example,  in the case of QCD, the fermion
 self energy damping at least as $1/p^2$ at ultraviolet momentum region,
 if we take a hard fermion
 mass to substitute self energy, we will over estimate its contribution
 to physics. To respect this momentum dependence of fermion self energy
in the theory ,
 in this paper, we develop a generalized proper time formulation to calculate
 fermion determinant with momentum dependent fermion self energy.

Since momentum dependent fermion self energy $\Sigma(p^2)$ in
coordinate space is represented by
$\Sigma(-\partial^2)\delta(x-y)$, the naive generalization of
lndet$(D+m)$ should be calculating lndet$[D+\Sigma(-\partial^2)]$.
We argue this is not suitable since original
 lndet$(D+m)$ is invariant under following local symmetry transformations,
 \begin{eqnarray}
 s(x)&\rightarrow& s'(x)=V(x)s(x)V^{\dagger}(x)\nonumber\\
 p(x)&\rightarrow& p'(x)=V(x)p(x)V^{\dagger}(x)\nonumber\\
 v_{\mu}(x)&\rightarrow& v_{\mu}'(x)=V(x)v_{\mu}(x)V^{\dagger}(x)
 +V(x)[i\partial_{\mu}V^{\dagger}(x)]\label{vtrans}\\
 a_{\mu}(x)&\rightarrow& a_{\mu}'(x)=V(x)a_{\mu}(x)V^{\dagger}(x)\;\;,\nonumber
  \end{eqnarray}
which leads to
\begin{eqnarray}
D_x\rightarrow D_x'&\equiv&
\partial\!\!\!/\;_x-iv\!\!\! /\;'(x)-ia\!\!\! /\;'(x)\gamma_5
-s'(x)+ip'(x)\gamma_5\nonumber\\
&=&V(x)[\partial\!\!\!/\;_x-iv\!\!\! /\;(x)-ia\!\!\! /\;(x)\gamma_5
-s(x)+ip(x)\gamma_5]V^{\dagger}(x)=V(x)D_xV^{\dagger}(x)
\end{eqnarray}
and
\begin{eqnarray}
{\rm lndet}(D+m)&\rightarrow& {\rm lndet}(D'+m)
 = {\rm lndet}(VDV^{\dagger}+m) =  {\rm lndet}[V(D+m)V^{\dagger}]
 \nonumber\\
 &=& {\rm lndet}(D+m)\;\;.
\end{eqnarray}
Here the invariance is due to the property of constant $m$ which leads
\begin{eqnarray}
V(x)mV^{\dagger}(x)=m\;\;.
\end{eqnarray}
If we change $m$ to $\Sigma(-\partial^2)$, above relation no
longer valid due to differential operator dependence of $\Sigma$,
\begin{eqnarray}
V(x)\Sigma(-\partial^2_x)V^{\dagger}(x)&=&
\Sigma[-V(x)\partial^2_xV^{\dagger}(x)]
=\Sigma\bigg[-\bigg(\partial_{\mu}+V(x)[\partial_{\mu}V^{\dagger}(x)]\bigg)^2
\bigg]\neq\Sigma(-\partial^2_x)\;\;.
\end{eqnarray}
To implement local symmetry (\ref{vtrans}) in our generalization,
instead of considering
$\Sigma(-\partial^2)$, we need to consider $\Sigma(-\overline{\nabla}^2)$ where
\begin{eqnarray}
\overline{\nabla}^{\mu}\equiv\partial^{\mu}-iv^{\mu}(x)\;\;,
\end{eqnarray}
where the bar over $\nabla_{\mu}$ is to specify the difference of present derivative with that introduced in (\ref{nabladef}).
Use (\ref{vtrans}), we find $\overline{\nabla}_{\mu}$  transform
as
\begin{eqnarray}
\overline{\nabla}^{\mu}_x\rightarrow\overline{\nabla}^{\mu\prime}_x
\equiv\partial^{\mu}_x-iv^{\mu\prime}(x)
=V(x)\overline{\nabla}^{\mu}_xV^{\dagger}(x)\;\;.
\end{eqnarray}
Then
\begin{eqnarray}
\Sigma(-\overline{\nabla}^2_x)&\rightarrow&
\Sigma(-\overline{\nabla}^{2\prime}_x)=
\Sigma[-V(x)\overline{\nabla}^2_xV^{\dagger}(x)]=
V(x)\Sigma(-\overline{\nabla}^2_x)V^{\dagger}(x)
\end{eqnarray}
and
\begin{eqnarray}
 &&{\rm lndet}[D+ \Sigma(\overline{\nabla}^2) ]\nonumber\\
 &&\rightarrow {\rm lndet}[D'+ \Sigma(\overline{\nabla}^{2\prime})]
   =  {\rm lndet}\bigg[V[D+ \Sigma(\overline{\nabla}^2)   ]V^{\dagger}\bigg]
 = {\rm lndet}[D+ \Sigma(\overline{\nabla}^2)]\;\;.
\end{eqnarray}
So, ${\rm lndet}[D+ \Sigma(\overline{\nabla}^2) ]$ is invariant under
transformation (\ref{vtrans}) and can be thought as correct
generalization of lndet$(D+m)$.  The generalized fermion determinant now is
\begin{eqnarray}
{\rm Re}\ln{\rm Det}[D+\Sigma(-\overline{\nabla}^2)]
&=&\frac{1}{2}{\rm Tr}\ln\bigg[[D^{\dagger}+\Sigma(-\overline{\nabla}^2)]
[D+\Sigma(-\overline{\nabla}^2)]\bigg]\nonumber\\
&=&-\frac{1}{2}\lim_{\Lambda\rightarrow\infty}
\int_{\frac{1}{\Lambda^2}}^{\infty}\frac{d\tau}{\tau}~
{\rm Tr}e^{-\tau[\overline{E}
-\nabla^2+\Sigma^2(-\overline{\nabla}^2)
+Jg(\overline{\nabla}^2)+g'(\overline{\nabla}^2)K
-d\!\!\! /\;\Sigma(-\overline{\nabla}^2)]}\label{lnSigma}
\end{eqnarray}
where
\begin{eqnarray}
\overline{E}-\nabla^2
+\Sigma^2(-\overline{\nabla}^2)
+Jg(\overline{\nabla}^2)+g'(\overline{\nabla}^2)K
-d\!\!\! /\;\Sigma(-\overline{\nabla}^2)
=[D^{\dagger}+\Sigma(-\overline{\nabla}^2)][D+\Sigma(-\overline{\nabla}^2)]
\end{eqnarray}
and
\begin{eqnarray}
&&[d\!\!\! /\;\Sigma(-\overline{\nabla}^2)]\equiv
\gamma^{\mu}[d_{\mu}\Sigma(-\overline{\nabla}^2)]=
\gamma^{\mu}\bigg(\partial_{\mu}\Sigma(-\overline{\nabla}^2)
-i[v_{\mu},\Sigma(-\overline{\nabla}^2)]\bigg)\nonumber\\
&&\overline{E}\equiv\frac{i}{4}[\gamma^{\mu},\gamma^{\nu}] {\cal
R}_{\mu\nu}+\gamma_{\mu}d^{\mu}(s+ip\gamma_5)
+i\gamma^{\mu}[a_{\mu}\gamma_5(s+ip\gamma_5)+(s+ip\gamma_5)a_{\mu}\gamma_5]
+s^2+p^2+[s,p]i\gamma_5\nonumber\\
&&g(x)=g'(x)\equiv\Sigma(-x)\hspace{1.5cm} J=-ia\!\!\!
/\;\gamma_5-s-ip\gamma_5\hspace{2cm} K=-ia\!\!\!
/\;\gamma_5-s+ip\gamma_5\;\;.\nonumber
\end{eqnarray}
For safety of further calculation, we limit the ultraviolet
behavior of $\Sigma(k^2)$ as
\begin{eqnarray}
\frac{\Sigma^2(k^2)}{k^2}\stackrel{k^2\rightarrow\infty}{--\rightarrow}0
\label{asymp}
\end{eqnarray}
The key now is to calculate
\begin{eqnarray}
&&{\rm Tr}e^{-\tau[\overline{E}
-\nabla^2+\Sigma^2(-\overline{\nabla}^2)
+Jg(\overline{\nabla}^2)+g'(\overline{\nabla}^2)K
-d\!\!\! /\;\Sigma(-\overline{\nabla}^2)]}\nonumber\\
&&=\int d^4x~{\rm tr}\langle x|e^{-\tau[\overline{E}
-\nabla^2+\Sigma^2(-\overline{\nabla}^2)
+Jg(\overline{\nabla}^2)+g'(\overline{\nabla}^2)K
-d\!\!\! /\;\Sigma(-\overline{\nabla}^2)]}|x\rangle
\end{eqnarray}
in which term $\langle x|e^{-\tau[\overline{E}
-\nabla^2+\Sigma^2(-\overline{\nabla}^2)
+Jg(\overline{\nabla}^2)+g'(\overline{\nabla}^2)K -d\!\!\!
/\;\Sigma(-\overline{\nabla}^2)]}|x\rangle$ is much more complex
than (\ref{Seely}), since except the constraint (\ref{asymp}),
the differential operator dependent function $\Sigma$ is still unknown.
\begin{eqnarray}
&&\langle x|e^{-\tau[\overline{E}
-\nabla^2+\Sigma^2(-\overline{\nabla}^2)
+Jg(\overline{\nabla}^2)+g'(\overline{\nabla}^2)K
-d\!\!\! /\;\Sigma(-\overline{\nabla}^2)]}|x\rangle\nonumber\\
&&=\int\frac{d^4p}{(2\pi)^4}\exp\bigg\{-\tau\bigg[\overline{E}(x)
-\nabla_x^2-2ip\cdot\nabla_x+p^2
+\Sigma^2(-\overline{\nabla}^2-2ip\cdot\overline{\nabla}_x+p^2)
\nonumber\\
&&\hspace{0.5cm}
+Jg(\overline{\nabla}_x^2+2ip\cdot\overline{\nabla}_x-p^2)
+g'(\overline{\nabla}_x^2+2ip\cdot\overline{\nabla}_x-p^2)K
-d\!\!\! /\;\Sigma(-\overline{\nabla}_x^2-2ip\cdot\overline{\nabla}_x+p^2)
\nonumber\\
&&\hspace{0.5cm}
-i[p\!\!\! /\;,\Sigma(-\overline{\nabla}_x^2-2ip\cdot\overline{\nabla}_x+p^2)]
\bigg]\bigg\}\;\;.\label{momexp}
\end{eqnarray}
Assign $\overline{E},J,K$ are order of $p$, we can
take low energy expansion for (\ref{momexp}). Substitute the result
of momentum expansion  into (\ref{lnSigma}), we finally get
\begin{eqnarray}
&&{\rm Re}\ln{\rm Det}[D+\Sigma(-\overline{\nabla}^2)]\nonumber\\
&&=\int d^4x{\rm  tr}_f\bigg[{\cal C}_0s+{\cal C}_1a^2+
 {\cal C}_2[d_{\mu}a^{\mu}]^2
 +{\cal C}_3(d^{\mu}a^{\nu}-d^{\nu}a^{\mu})(d_{\mu}a_{\nu}-d_{\nu}a_{\mu})
+{\cal C}_4a^4\label{detresult}\\ &&\hspace{0.5cm}
 +{\cal C}_5a^{\mu}a^{\nu}a_{\mu}a_{\nu} +{\cal C}_6s^2+{\cal C}_7p^2
+{\cal C}_8sa^2 +{\cal C}_9V^{\mu\nu}V_{\mu\nu}+{\cal
C}_{10}V^{\mu\nu}a_{\mu}a_{\nu} +{\cal
C}_{11}pd_{\mu}a^{\mu}\bigg]+O(p^6)\nonumber
\end{eqnarray}
with coefficients related to $\Sigma$ by
\begin{eqnarray}
{\cal C}_0&=&-4\int d\tilde{k} \Sigma_kX_k\nonumber\\
 {\cal C}_1&=&2\int d\tilde{k}\bigg[
(-2\Sigma^2_k+k^2\Sigma_k\Sigma'_k)X_k^2+(-2\Sigma^2_k+k^2\Sigma_k\Sigma'_k)
\frac{X_k}{\Lambda^2}\bigg]\nonumber\\
 {\cal C}_2&=&-2\int d\tilde{k}\bigg[2A_kX_k^3
+2A_k\frac{X_k^2}{\Lambda^2}+A_k\frac{X_k}{\Lambda^4}
+\frac{k^2}{2}\Sigma'^2_k\frac{X_k}{\Lambda^2}+\frac{k^2}{2}\Sigma'^2_kX_k^2
\bigg]\nonumber\\
 {\cal C}_3&=&-\int d\tilde{k}\bigg[2B_kX_k^3
 +2B_k\frac{X_k^2}{\Lambda^2}+B_k\frac{X_k}{\Lambda^4}
+\frac{k^2}{2}\Sigma^{\prime 2}_k\frac{X_k}{\Lambda^2}
+\frac{k^2}{2}\Sigma^{\prime 2}_kX_k^2\bigg]\nonumber\\
 {\cal C}_4&=&2\int d\tilde{k}\bigg[
(\frac{4\Sigma^4_k}{3}+\frac{2k^2\Sigma^2_k}{3}+\frac{k^4}{18})(
 6X_k^4+\frac{6X_k^3}{\Lambda^2}+\frac{3X_k^2}{\Lambda^4}
+\frac{X_k}{\Lambda^6})-(4\Sigma^2_k+\frac{k^2}{2})(
2X_k^3\nonumber\\
&&+\frac{2X_k^2}{\Lambda^2}+\frac{X_k}{\Lambda^4})
+\frac{X_k}{\Lambda^2}+X_k^2\bigg]\nonumber\\
 {\cal C}_5&=&\int d\tilde{k}\bigg[
(\frac{-4\Sigma^4_k}{3}-\frac{2k^2\Sigma^2_k}{3}+\frac{k^4}{18})(
6X_k^4+\frac{6X_k^3}{\Lambda^2}+\frac{3X_k^2}{\Lambda^4}
+\frac{X_k}{\Lambda^6})
+4\Sigma^2_k(2X_k^3+\frac{2X_k^2}{\Lambda^2}\nonumber\\
&&+\frac{X_k}{\Lambda^4})-\frac{X_k}{\Lambda^2}-X_k^2\bigg]\nonumber\\
{\cal C}_6&=&2\int
d\tilde{k}\bigg[(3\Sigma^2_k-2k^2\Sigma_k\Sigma'_k)X_k^2
+[2\Sigma^2_k-k^2(1+2\Sigma_k\Sigma'_k)]\frac{X_k}{\Lambda^2}\bigg]\nonumber\\
{\cal C}_7&=&2\int
d\tilde{k}\bigg[(\Sigma^2_k-2k^2\Sigma_k\Sigma'_k)X_k^2
-k^2(1+2\Sigma_k\Sigma'_k)\frac{X_k}{\Lambda^2}\bigg]\nonumber\\
{\cal C}_8&=&-4\int d\tilde{k}\bigg[(4\Sigma^3_k+k^2\Sigma_k)X_k^3
-(4\Sigma^3_k+k^2\Sigma_k)\frac{X_k}{\Lambda^2}
+(2\Sigma^3_k+\frac{1}{2}k^2\Sigma_k)\frac{X_k}{\Lambda^4}
-3\Sigma_k\frac{X_k}{\Lambda^2}\nonumber\\
&&-3\Sigma_kX_k^2\bigg]\nonumber\\ {\cal C}_9&=&-\int
d\tilde{k}\bigg[
(\frac{1}{2}k^2\Sigma'_k\Sigma''_k+\frac{1}{6}k^2\Sigma_k\Sigma'''_k)X_k
+(-C_k+D_k)\frac{X_k}{\Lambda^2}
-(C_k-D_k)X_k^2+2E_kX_k^3\nonumber\\
&&+2E_k\frac{X_k^2}{\Lambda^2}
-E_k\frac{X_k^2}{\Lambda^4}\bigg]\nonumber\\
 {\cal C}_{10}&=&-4i\int d\tilde{k}\bigg[
2F_kX_k^3+2F_k\frac{X_k^2}{\Lambda^2}
+F_k\frac{X_k}{\Lambda^4}
+\frac{k^2}{2}\Sigma_k^{\prime 2}\frac{X_k}{\Lambda^2}
+\frac{k^2}{2}\Sigma^{\prime 2}_kX_k^2\bigg]
 \nonumber\\
{\cal C}_{11}&=&4\int d\tilde{k}\bigg[
(\Sigma_k-\frac{1}{2}k^2\Sigma'_k)\frac{X_k}{\Lambda^2}
+(\Sigma_k-\frac{1}{2}k^2\Sigma'_k)X_k^2\bigg]\label{Kresult}
\end{eqnarray}
where
\begin{eqnarray}
\int d\tilde{k}&\equiv&
N_c\int\frac{d^4k}{(2\pi)^4}e^{-\frac{k^2+\Sigma^2(k^2)}{\Lambda^2}}
\label{measure}\\
\Sigma_k&\equiv&\Sigma(k^2)\hspace{2cm}
X_k\equiv\frac{1}{k^2+\Sigma^2(k^2)}
\end{eqnarray}
and $A_k,B_k,C_k,D_k,E_k,F_k$ are given in appendix \ref{Coef}.
We see that asymptotic behavior of $\Sigma(k^2)$ (\ref{asymp}) insure
factor $e^{-\frac{k^2+\Sigma^2(k^2)}{\Lambda^2}}$ appeared in integration
 measure (\ref{measure}) is an ultraviolet suppression factor which will keep
 our momentum integration be convergent.

(\ref{detresult}) and (\ref{Kresult}) are our final result for the real part of
fermion determinant with presence of dynamical quark self energy. The result in
this paper only given up to order of $p^4$, one can easily generalize the
calculation to higher orders of the momentum expansion. As a self check of theory, take $\Sigma(k^2)$ be constant $m$, in the limit of
$\Lambda^2\rightarrow\infty$, the momentum integration in (\ref{Kresult})
can be finished, the result gives
\begin{eqnarray}
&&{\cal C}_0\stackrel{\Sigma=m}{--\rightarrow}
-\frac{N_c}{4\pi^2}m[\Lambda^2+m^2(ln\frac{m^2}{\Lambda^2}+\gamma-1)]
\nonumber\\ &&{\cal C}_1\stackrel{\Sigma=m}{--\rightarrow}
\frac{N_c}{4\pi^2}m^2(ln\frac{m^2}{\Lambda^2}+\gamma)\nonumber\\
&&{\cal C}_2 \stackrel{\Sigma=m}{--\rightarrow}
\frac{N_c}{24\pi^2}\nonumber\\ && {\cal
C}_3\stackrel{\Sigma=m}{--\rightarrow}
\frac{N_c}{48\pi^2}(ln\frac{m^2}{\Lambda^2}+\gamma+1)\nonumber\\
&&{\cal C}_4\stackrel{\Sigma=m}{--\rightarrow}
\frac{N_c}{24\pi^2}(ln\frac{m^2}{\Lambda^2}+\gamma+4)\nonumber\\
&&{\cal C}_5\stackrel{\Sigma=m}{--\rightarrow}
-\frac{N_c}{24\pi^2}(ln\frac{m^2}{\Lambda^2}+\gamma+2)\nonumber\\
&&{\cal C}_6\stackrel{\Sigma=m}{--\rightarrow}
\frac{N_c}{8\pi^2}\bigg[\Lambda^2+m^2(3ln\frac{m^2}{\Lambda^2}+3\gamma-1)
\bigg]\nonumber\\ &&{\cal C}_7\stackrel{\Sigma=m}{--\rightarrow}
\frac{N_c}{8\pi^2}\bigg[\Lambda^2+m^2(ln\frac{m^2}{\Lambda^2}+\gamma-1)\bigg]
\nonumber\\ &&{\cal C}_8\stackrel{\Sigma=m}{--\rightarrow}
-\frac{N_c}{2\pi^2}m(ln\frac{m^2} {\Lambda^2}+\gamma+1)
\nonumber\\ &&{\cal C}_9\stackrel{\Sigma=m}{--\rightarrow}
\frac{N_c}{48\pi^2}(ln\frac{m^2} {\Lambda^2}+\gamma)\nonumber\\
&&{\cal C}_{10}\stackrel{\Sigma=m}{--\rightarrow}
-\frac{iN_c}{12\pi^2}(ln\frac{m^2}{\Lambda^2}+\gamma+2)\nonumber\\
&&{\cal C}_{11}\stackrel{\Sigma=m}{--\rightarrow}
-\frac{N_c}{4\pi^2}m(ln\frac{m^2}{\Lambda^2}+\gamma)\;\;.
\end{eqnarray}
Substitute them into (\ref{detresult}),  we reproduce
original result (\ref{exp}).

  In summary, we have generalized the traditional proper time method for calculation  of fermion determinant to include dynamical chiral symmetry breaking caused by  momentum dependent  fermion self energy.
   The physical application of this formulation will be discussed elsewhere.

\section*{Acknowledgments}

This work was  supported by fundamental research grant of Tsinghua University.


\appendix

\section{Coefficients definitions}\label{Coef}

\begin{eqnarray}
A_k&=&\frac{2}{3}k^2\Sigma_k\Sigma'_k(-1-2\Sigma_k\Sigma'_k)-\frac{1}{3}
\Sigma^2_k(-1-2\Sigma_k\Sigma'_k)
+\frac{1}{3}k^2\Sigma^2_k(\Sigma^{\prime 2}_k+
\Sigma_k\Sigma''_k)\nonumber\\
&&+\frac{1}{6}k^4(\Sigma^{\prime 2}_k+\Sigma_k\Sigma''_k)
\nonumber\\
B_k&=&\frac{2}{3}k^2\Sigma_k\Sigma'_k(-1-2\Sigma_k\Sigma'_k)-\frac{1}{3}
\Sigma^2_k(-1-2\Sigma_k\Sigma'_k)+\frac{1}{3}k^2\Sigma^2_k(\Sigma^{\prime 2}_k
+\Sigma_k\Sigma''_k)\nonumber\\
&&+\frac{1}{18}k^4(\Sigma^{\prime 2}_k+\Sigma_k\Sigma''_k)
+\frac{1}{6}k^2(-1-2\Sigma_k\Sigma'_k)\nonumber\\
C_k&=&\frac{1}{3}-\frac{1}{3}\Sigma_k\Sigma'_k
+\frac{1}{2}k^2\Sigma^{\prime 2}_k\nonumber\\
D_k&=&-\frac{1}{2}k^2\Sigma^{\prime 2}_k+\frac{1}{3}k^2\Sigma_k\Sigma''_k
(-1-2\Sigma_k\Sigma'_k)+\frac{2}{9}k^4\Sigma'_k\Sigma''_k
(1+2\Sigma_k\Sigma'_k)]
-\frac{2}{9}k^4\Sigma^{\prime 2}_k(-\Sigma^{\prime 2}_k-\Sigma_k\Sigma''_k)
\nonumber\\
&&+\frac{1}{3}k^2\Sigma_k\Sigma'_k(-\Sigma^{\prime 2}_k-\Sigma_k\Sigma''_k)
\nonumber\\
E_k&=&\frac{1}{6}k^2\Sigma_k\Sigma'_k(-1-2\Sigma_k\Sigma'_k)^2
+\frac{1}{9}k^4\Sigma^{\prime 2}_k(1+2\Sigma_k\Sigma'_k)^2\nonumber\\
F_k&=&\frac{4}{3}k^2\Sigma_k\Sigma'_k-
\frac{4}{3}k^2(\Sigma_k\Sigma'_k)^2-\frac{2}{3}\Sigma^2_k
+\frac{2}{3}\Sigma^3_k\Sigma'_k
+\frac{1}{3}k^2\Sigma^2_k(\Sigma^{\prime 2}_k+\Sigma_k\Sigma''_k)
+\frac{1}{9}k^4(\Sigma^{\prime 2}_k+\Sigma_k\Sigma''_k)\nonumber\\
&&+\frac{1}{3}k^2(-1-2\Sigma_k\Sigma'_k)+\frac{1}{2}k^2\nonumber
\end{eqnarray}

\end{document}